\documentclass[12pt]{article}
\usepackage{amsfonts}
\usepackage{bm}
\usepackage{amsmath}
\usepackage{amssymb}
\usepackage{epsfig}
\usepackage[mathscr]{euscript}
\usepackage{hyperref}

\usepackage{slashed}

\newcommand{\bmat}{\left(\begin{array}}
\newcommand{\emat}{\end{array}\right)}

\def\gtrsim{\mathrel{\raise.3ex\hbox{$>$\kern-.75em\lower1ex\hbox{$\sim$}}}}

\def\-{\hphantom{-}}
\def\ov{\overline}
\def\un{\underline}
\def\s2{\frac{1}{\sqrt2}}

\def\mg{m_{3/2}}
\def\mg2{m^2_{3/2}}

\def\Dsl{\,\raise.15ex\hbox{/}\mkern-13.5mu D} 

\def\be{\begin{equation}}
\def\ee{\end{equation}}
\def\bea{\begin{eqnarray}}
\def\eea{\end{eqnarray}}

\newcommand{\nn}{\nonumber}

\begin{document}

\pagestyle{plain}

\makeatletter
\@addtoreset{equation}{section}
\makeatother
\renewcommand{\theequation}{\thesection.\arabic{equation}}
\pagestyle{empty}
\begin{center}
\ \

\vskip .5cm
\LARGE{\LARGE\bf ${\cal N}=1$ supersymmetry and Non-Riemannian Double Field Theory
\\[10mm]}
\vskip 0.3cm

\large{Eric Lescano
 \\[6mm]}
 
 {\small  Division of Theoretical Physics, Rudjer Boskovic Institute \\ [.01 cm]}
{\small\it Bijenicka 54, 10000 Zagreb, Croatia\\ [.2 cm]}

{\small \verb"elescano@irb.hr"}\\[1cm]

\small{\bf Abstract} \\[0.5cm]\end{center}
We construct the ${\cal N}=1$ supersymmetric extension of Double Field Theory for Riemannian and the non-Riemannian in a unified approach. The inclusion of fermions in the double geometry force us to use the generalized frame formalism to construct the generalized flux components for 
these geometries. We focus on the most general prescription required to get the $D=10$ minimal supergravity model. We study how to consistently avoid the gauge fixing procedure of the double Lorentz symmetry when $n=\bar n$ and $0\le n \le 5$, which gives rise to a bigravity structure (pair of vielbeins producing the same non-Riemannian degrees of freedom). As an example we show how to to include fermionic degrees of freedom in the type I torsional Newton-Cartan (TNC) theory ($n=\bar n=1$) which is related to Carrollian geometries and stringy Newton Cartan through duality rotations and/or null reductions/uplifts.

\newpage

\setcounter{page}{1}
\pagestyle{plain}
\renewcommand{\thefootnote}{\arabic{footnote}}
\setcounter{footnote}{0}


\section{Introduction}

Double Field Theory (DFT) \cite{Siegel,DFT} is typically understood as a convenient rewriting of the low energy limit of string theory in terms of $O(D,D)$ multiplets. All the massless fields of the universal NS-NS sector can be regrouped in terms of a generalized metric, ${\cal H}_{M N}$ ($M,N=0 \dots 2D-1$) and a generalized dilaton $d$, while a variational principle can be defined in terms of these fields in a double space. Inspecting the most general parametrization of the former fields the authors in \cite{PM} showed that both Riemannian and non-Riemannian geometries can be rewriting in this framework. Moreover, they classified all the possible scenarios using two integers $(n,\bar n)$, where $n + \bar n=D$ and $n=\bar n=0$ corresponds to the Riemmanian case. 

Effective non-Riemannian geometries are backgrounds compatible with string theory \cite{NRST} and, interestingly enough, these backgrounds are related to Riemannian ones by T-duality transformations (or a null reduction/uplift) \cite{Null}. In this sense, the formulation introduced in \cite{PM} is a natural proposal for unifying all these scenarios in a duality invariant way.  The ${\cal N}=1$ supersymmetric extension of DFT was introduced in \cite{SDFT1} up to two fermions and in \cite{SDFT2} considering the full four fermions interactions. This extension requires the generalized frame/flux formalism of DFT \cite{FF} because of the presence of fermionic degrees of freedom and a Clifford algebra at the level of the double geometry. 

The main goal of this work is to implement the previous extension for Riemannian and non-Riemannian scenarios in a unified approach. Our main results are
\begin{itemize}
\item The construction of the generalized flux formalism for non-Riemannian DFT, in agreement with the generalized metric formulation.
\item The study of the gauge fixing procedure of the double Lorentz transformations for these scenarios which give rise to a bigravity structure for the degrees of freedom of the non-Riemannian supergravity.
\item The computation of the general action and (super) symmetry transformations for the non-Riemannian fields.

\item The ${\cal N}=1$ supersymmetric extension of TNC geometry coming from DFT.
\end{itemize}

In order to achieve the previous results we use the generalized frame parametrization introduced in \cite{PM} in order to construct the different components of the generalized fluxes that are required by the DFT action principle. Particularly, only $n=\bar n$ scenarios are compatible with the double Lorentz symmetry and therefore with supersymmetry as firstly noticed in \cite{PM} (see also the last section of \cite{nonPRL}). We construct the most general ${\cal N}=1$ extension to the DFT action principle in order to reproduce the minimal $D=10$ supergravity model and we read the transformation rules for the degrees of freedom of the non-Riemannian supergravity in agreement with the DFT construction. 

The bosonic part of this framework was given in \cite{NRaction} considering a generalized metric formulation, which is equivalent to the generalized flux formalism here presented. The only difference is that the latter forces $n=\bar n$ while the inclusion of supersymmetry fixes the dimension of the space-time from the very beginning. Therefore we work in $D=10$, while extensions to other non-Riemannian supergravities should be easily extracted from our model considering the proper Clifford algebra and fermionic relations. We will not assume a factorization for the geometry, but see \cite{KK} for a bosonic treatment between a Riemannian formulation on the external space and a non-Riemannian one on the internal space.

In this work we also show how this generic supersymmetric extension can be used in order to include fermionic degrees of freedom in the Torsional Newton-Cartan (TNC) gravity. This is a convenient example because it was explored at the (supersymmetric) worldsheet level in \cite{Blair} and it is a geometry related to Carrollian and stringy Newton-Cartan using T-duality and/or null reduction/uplift (See \cite{NRaction} for the specific transformations between TNC and Carrollian/stringy Newton-Cartan). In the last part of the work we discuss potential continuations for this line of investigation. Particularly we discuss about testing this model using perturbative theory and to extend the proposal to be able to describe the ${\cal N}=1$ supersymmetric extension for type II TNC.

\section{Flux formulation for Non-Riemannian Double Field Theory}

The construction of the flux formalism \cite{FF} of the non-Riemannian version of DFT \footnote{See \cite{ReviewDFT} for reviews of the topic.} can be performed considering four steps \footnote{Here we mimic the presentation of \cite{LectureSDC}, but for the non-Riemannian case. These lectures are also useful as an introduction to DFT.}. We start by describing the double geometry and the fundamental fields of the theory, with their non-relativistic parametrization. Then we inspect the symmetry rule transformations and finally we discuss about the action principle. Our conventions for the indices are as follows: $O(D,D)$ indices are given by $M,N,\dots=0,\dots,2D-1$ where $M=({}_{\dot \mu}, {}^{\dot \mu})$. Their flat indices (double Lorentz indices) are $A=(\underline A, \overline A)$, where $\underline A=\underline{\dot a}$ and $\overline A=\dot{\overline a}$. Space-time indices are $\dot \mu = 0, \dots, d-1$.

\begin{enumerate}
\item {\bf Double geometry}: We consider a double geometry with coordinates $X^{M}$. We equip a group invariant metric $\eta_{MN}$, 
\bea
{\eta}_{MN}  =  \left(\begin{matrix}0&\delta_{\dot \mu}^{\dot \nu}\\ 
\delta^{\dot \mu}_{\dot \nu}&0 \end{matrix}\right) \, ,  \label{etaintro}
\eea
(and its inverse $\eta^{MN}$) which is used to lower (and raise) double curved indices.  On each point of the double space we consider a double tangent space, so we are able to define flat vectors $V^A$. Then we consider two additional invariant and flat metrics $\eta_{A B}$ and ${\cal H}_{A B}$. The former is used to lower flat indices and both of them are used to construct the following flat projectors,
\bea
P_{AB} = \frac{1}{2}\left(\eta_{AB} - {\cal H}_{AB}\right) =  \left(\begin{matrix} P^{\underline {\dot a} \underline{\dot b}} & 0 \\ 0 & 0 \end{matrix}\right) \, , 
\label{etaflatupproj}
\eea
\bea
\ov{P}_{AB} = \frac{1}{2}\left(\eta_{AB} + {\cal H}_{AB}\right) = \left(\begin{matrix} 0 & 0 \\ 0 & \ov P^{\bar {\dot a} \bar{\dot b}}  \end{matrix}\right) \, ,
\eea
which satisfy
\bea
&{\overline{P}}_{{A B}} {\overline{P}}^{ B}{}_{C}={\overline{P}}_{{A C}}\, , &\quad {P}_{{A B}} {P}^{B}{}_{C}={P}_{{A C}}, \nn\\
&{P}_{{A  B}}{\overline{P}}^{B}{}_{ C} = {\overline{P}}_{ {A B}}  {P}^{ B}{}_{C} = 0\, ,  &\quad {\overline{P}}_{{AB}} + {P}_{{A B}} = \eta_{{A B}}\,.
\label{proy}
\eea
The invariant projectors of DFT are parametrized in the following way

\bea
P_{\underline{\dot a \dot b}}= \left(\begin{matrix} -\eta_{\underline a \underline b} & 0 & 0 \\ 0 & -\delta_{i j} & 0 \\ 0 & 0 & \delta_{i j} \end{matrix}\right) , \quad \overline P_{\overline{\dot a \dot b}}= \left(\begin{matrix} \eta_{\bar a \bar b}  & 0 & 0 \\ 0 & - \delta_{\bar i \bar j} & 0 \\ 0 & 0 & \delta_{\bar i \bar j} \end{matrix}\right)  \, . 
\eea

Thus, $P$ and $\bar P$ project onto complementary orthogonal subspaces and any arbitrary flat vector can be written as
\bea
V^{A} = P^{A}{}_{B} V^{B} + \ov P^{A}{}_{B} V^{B} = V^{\un A} + V^{\ov A} \, .
\label{projflat}
\eea

\item{\bf Fundamental fields:} The fundamental fields of DFT are a generalized frame $E_{MA}(X)=E_{MA}$ and a generalized dilaton $d(X)=d$. The former is equivalent to a vielbein for this double geometry and for non-Riemannian backgrounds its parametrization is
\be
E^{M}{}_{ A}  =\left(\begin{matrix}{ E}_{\dot \mu \underline{\dot a}}&  { E}_{}^{\dot \mu }{}_{\underline{\dot a}} \\ 
E_{\dot \mu \overline{\dot a}}& E_{}^{\dot \mu}{}_{\overline{\dot a}} \end{matrix}\right) \ = \
\frac{1}{\sqrt{2}}\left(\begin{matrix}- k_{\dot \mu \dot{\un a}}-B_{\dot \rho \dot \mu} {h}^{\dot \rho }{}_{\dot{\un a}} &  h^{\dot \mu }{}_{\dot{\un a}}  \\ 
\bar k_{\dot \mu \ov{\dot a}}-B_{\dot \rho \dot \mu}{} \bar h^{\dot \rho }{}_{\ov{\dot a}}& \bar h^{\dot \mu}{}_{\ov{\dot a}} \end{matrix}\right)  \, ,
\label{HKparam}
\ee
where
\bea
k_{\dot \mu \underline{\dot a}} = (k_{\dot \mu \underline{a}}, x_{\dot \mu},x_{\dot \mu})\, , \quad \bar k_{\dot \mu \overline{\dot a}} = (\bar k_{\dot \mu \overline{a}}, \bar x_{\dot \mu}, \bar x_{\dot \mu}) \, , \\
h^{\dot \mu}{}_{\underline{\dot a}} = (h^{\dot \mu}{}_{\underline{a}}, y^{\dot \mu}, y^{\dot \mu})\, , \quad \bar h^{\dot \mu}{}_{ \overline{\dot a}} = (\bar h^{\dot \mu}{}_{ \overline{a}}, \bar y^{\dot \mu}, \bar y^{\dot \mu}) \, ,
\eea
and 
\bea
\label{sq1}
\bar k_{\dot \mu}{}^{\ov{a}} \bar k_{\dot \nu \ov{a}} = k_{\dot \mu}{}^{\un{a}} k_{\dot \nu \un{a}} & = & K_{\dot \mu \dot \nu} \, , \\
\bar h^{\dot \mu}{}_{\ov{a}} \bar h^{\dot \nu \ov{a}} = h^{\dot \mu}{}_{\un{a}} h^{\dot \nu \un{a}} & = & H^{\dot \mu \dot \nu} \, .
\label{sq2}
\eea
In the previous expresion $K_{\dot \mu \dot \nu}$ and $H^{\dot \mu \dot \nu}$ are two symmetric tensors whose kernels are spanned by $\Big\{ x_{\dot \mu}, \bar x_{\dot \mu} \Big\}$ and $\Big\{ y^{\dot \mu}, \bar y^{\dot \mu} \Big\}$, respectively,
\bea
H^{\dot \mu \dot \nu} x_{\dot \nu} = H^{\dot \mu \dot \nu} \bar x_{\dot \nu} = K_{\dot \mu \dot \nu} y^{\dot \nu} = K_{\dot \mu \dot \nu} \bar y^{\dot \nu} = 0 \, ,
\eea
while $B_{\dot \mu \dot \nu}$ is identified the ordinary B-field. 

We define the following completeness relations, 
\bea
\delta^{\dot \mu}_{\dot \nu} & = & H^{\dot \mu \dot \rho} K_{\dot \nu \dot \rho} + y^{\dot \mu} x_{\dot\nu} + \bar y^{\dot \mu} \bar x_{\dot \nu} \, , \\
 \delta_{\dot \mu}^{\dot \nu} + x_{\dot \mu} y^{\dot \nu} - \bar x_{\dot \mu} \bar y^{\dot \nu} & = &\bar k_{\dot \mu \ov{\dot a}} \bar h^{\dot \nu \ov{\dot a}}  \, , \\
  \delta_{\dot \mu}^{\dot \nu} - x_{\dot \mu} y^{\dot \nu} + \bar x_{\dot \mu} \bar y^{\dot \nu} & = & k_{\dot \mu \un{\dot a}} h^{\dot \nu \un{\dot a}} \, , \\
 \delta_{\un{\dot a}}^{\un{\dot b}}  & = & k_{\dot \mu \un{\dot a}} h^{\dot \mu \un{\dot b}} \, , \\ 
 \delta_{\ov{\dot a}}^{\ov{\dot b}}  & = & \bar k_{\dot \mu \ov{\dot a}} \bar h^{\dot \mu \ov{\dot b}} \, .
\eea

With the generalized frame it is possible to construct the generalized metric and the invariant metric,
\bea
E_{M A} {\cal H}^{A B} E_{N B} & = & {\cal H}_{M N} \, , \nn \\
E_{M A} \eta^{A B} E_{N B} & = & \eta_{M N} \, .
\label{metricDFT}
\eea
The conventions for this paper were chosen in order to the generalized metric coincides with the one in \cite{NRaction},
\begin{align}\label{parametrization}
	\mathcal{H}_{MN} &= \begin{pmatrix}
	 K_{\mu\nu} - {B}_{\mu \rho} H^{\rho \sigma} {B}_{\sigma \nu} + 2 x^i_{(\mu} {B}_{\nu)\rho} y^\rho_i - 2 \bar x^{\bar{\imath}}_{(\mu} {B}_{\nu) \rho}\bar y^\rho_{\bar{\imath}} 	 & & - H^{\nu \rho} {B}_{\rho \mu} + y^\nu_i x^i_\mu - \bar y^\nu_{\bar{\imath}} \bar x^{\bar{\imath}}_{\mu} \\
		 - H^{\mu \rho} {B}_{\rho \nu} + y^\nu_i x^i_\mu - \bar y^\nu_{\bar{\imath}} \bar x^{\bar{\imath}}_{\mu} & & H^{\mu \nu} \, 
	\end{pmatrix} \, ,
\end{align}
which can be factorized as follows \cite{PC}
\begin{align}\label{conjugation}
	\mathcal{H}_{MN}= \begin{pmatrix}
		1 & {B} \\
		0 & 1
	\end{pmatrix} \begin{pmatrix}
		K& Z \\
		Z^T & H  
	\end{pmatrix} \begin{pmatrix}
		1 &0 \\
		 -  {B} & 1
	\end{pmatrix}
\end{align}
where we defined
\be
Z^\mu_{\ \nu} \equiv y^\mu_i x_\nu^i - \bar y^\mu_{\bar{\imath}} \bar  x_\nu^{\bar{\imath}}.
\ee
 Finally, with the help of (\ref{metricDFT}) one can define curved projectors,
\bea
P_{MN} = \frac{1}{2}\left(\eta_{MN} - {\cal  H}_{MN}\right) \, , \ \ \ \ \
\ov{P}_{MN} = \frac{1}{2}\left(\eta_{MN} + {\cal H}_{MN}\right)\ .
\label{proyc}
\eea
which satisfy
\bea
&{\overline{P}}_{{M Q}} {\overline{P}}^{ Q}{}_{ N}={\overline{P}}_{{M N}}\, , &\quad {P}_{{M Q}} {P}^{Q}{}_{ N}={P}_{{M N}}, \nn\\
&{P}_{{M  Q}}{\overline{P}}^{Q}{}_{ N} = {\overline{P}}_{ {M Q}}  {P}^{ Q}{}_{ N} = 0\, ,  &\quad {\overline{P}}_{{MN}} + {P}_{{M N}} = \eta_{{M N}}\, .
\label{curveprojrel}
\eea

\item {\bf Symmetries:} DFT is a T-duality invariant formulation, which can describe both Riemannian and non-Riemannian backgrounds. Consequently, all the fields and parameters are written in representations of the duality group and duality invariance is always guaranteed. 

We can also define generalized diffeomorphisms. These are infinitesimal transformations acting on a generic double vector $V^M$ through a generalized Lie derivative, \textit{i.e.},
\bea
\delta_{\hat \xi} V^M = {\mathcal L}_{\hat \xi} V^M =\hat \xi^N \partial_{N} V^M + (\partial^M \hat \xi_{P} - \partial_P \hat \xi^M) V^{P} + \omega \partial_{N} \hat \xi^N V^{M}\, .
\eea

In the previous expression we consider a generic parameter $\hat \xi^{M}$ and a density weight factor $\omega$. The closure of the generalized diffeomorphism transformations, 
\bea
\Big[\delta_{\hat \xi_1},\delta_{\hat \xi_2} \Big] V^{M}{}_{N} = \delta_{\hat \xi_{21}} V^{M}{}_{N} \,,
\eea
is provided by the C-bracket,
\bea
\hat \xi^{M}_{12} = \hat \xi^{P}_{1} \frac{\partial \hat \xi^{M}_{2}}{\partial X^{P}} - \frac12 \hat \xi^{P}_{1} \frac{\partial \hat \xi_{2P}}{\partial X_{M}} - (1 \leftrightarrow 2) \, .
\label{Cbra}
\eea
Moreover the closure is satisfied imposing the strong constraint
\bea
\partial_{M} \star \partial^{M} \star & = & 0 \, , \\
\partial_{M}(\partial^{M}) & = & 0 \, .
\eea
These constraints are solved using $\partial_{M}=(0,\partial_{\mu})$, which means that the generalized fields and parameters do not depend on the dual coordinates. On the other hand, the DFT Jacobiator is not trivial (but it is given by a trivial parameter) and the algebraic structure of DFT is given by an $L_{\infty}$ algebra with a non-trivial $l_3$ product \cite{HZ}, which measures the failure of the Jacobi identity in the double geometry. 

The generalized metric transforms as a tensor with $\omega=0$ with respect to generalized diffeomorphisms and, when non-Riemannian backgrounds are considered, one obtains that $H^{\dot \mu \dot \nu}$ and $K_{\dot \mu \dot \nu}$ transforms as tensors under ordinary diffeomorphisms, 
\bea
\delta_{\xi} H^{\dot \mu \dot \nu} = L_{\xi} H^{\dot \mu \dot \nu}, \quad \delta_{\xi} K_{\dot \mu \dot \nu} = L_{\xi} K_{\dot \mu \dot \nu} \, .  
\eea
As usually the B-field receives a gauge transformation,
\bea
\delta B_{\dot \mu \dot \nu} = L_{\xi} B_{\dot \mu \dot \nu} + 2 \partial_{[\dot \mu} \zeta_{\dot \nu]} \, ,
\eea
where $\xi_{M} = (\zeta_{\dot \mu},\xi^{\dot \mu})$.

The generalized metric \eqref{parametrization} is invariant under $GL(n)\times GL(\bar n)$ rotations 
\begin{align}
	\left(x^i_\mu, y^\mu_i , \bar x^{\bar{\imath}}_\mu, \bar y^\nu_{\bar{\imath}} \right) \rightarrow \left( x^j_\mu R^i_j, \left(R^{-1} \right)_i^{\hphantom{i}j} y^\nu_j, \bar x^{\bar{\jmath}}_\mu \bar R_{\bar{\jmath}}^{\hphantom{\bar{\jmath}} \bar{\imath}} , \left(\bar R ^{-1}\right)_{\bar{\imath}}^{\hphantom{\bar{\imath}} \bar{\jmath}} \bar y^\nu_j \right)\, ,
\end{align}
and under the generalized shift symmetry  
\begin{align} \label{shiftDFT}
\begin{split}
\left(y^\mu_i\right)' &= y_i^\mu +V^{\mu}_ i\\
\left(\bar y^\mu_{\bar{\imath}}\right)' &= 	\bar y_{\bar{\imath}}^\mu + \bar V^{\mu}_{\bar{\imath}} \\
\left(K_{\mu\nu}\right)' &= K_{\mu\nu}-2x^i_{(\mu}K_{\nu)\rho}  V^\rho_{i}-2\bar x^{\bar{\imath}}_{(\mu}K_{\nu)\rho}\bar V^\rho _{\bar{\imath}} +\left(x_\mu^i  V_{\rho i} +\bar x^{\bar{\imath}}_\mu \bar V_{\rho \bar{\imath}}  \right) \left(x^i_\nu  V^\rho_i +\bar x^{\bar{\imath}}_\nu\bar V^\rho_{\bar{\imath}} \right)  \\
\left({B}_{\mu\nu}\right)' &= {B}_{\mu\nu}-2x^i_{[\mu}  V_{\nu]i}+2\bar x^{\bar{\imath}}_{[\mu}\bar  V_{\nu]\bar{\imath}} + 2x^i_{[\mu}\bar x^{\bar{\jmath}}_{\nu]}\left(y_i^\rho \bar V_{\rho \bar{\jmath}} +\bar y_{\bar{\jmath}}^\rho   V_{\rho i} +V_{\rho i} \bar V^\rho_{\bar{\jmath}} \right) \nn
\end{split}
\end{align}
with $V_{\mu i}$ and $\bar V_{\mu \bar{\imath}}$ being the transformation parameters and we defined $V^\mu_{i}\equiv H^{\mu\rho}V_{\rho i}, \bar V^\mu_{\bar{\imath}}\equiv H^{\mu\rho}\bar V_{\rho \bar{\imath}}$.

Another symmetry of DFT are double Lorentz transformations, which act as
\bea
\delta_{\Lambda} V^{A} = V^{B} \Lambda_{B}{}^{A} \, ,
\eea
on an arbitrary flat vector $V^A$. Demanding $\delta_{\Lambda} \eta_{AB}=0$ we have $\Lambda_{AB}=-\Lambda_{BA}$. Moreover, using the decomposition/notation (\ref{projflat}), the condition $\delta_{\Lambda}{\cal H}_{AB}=0$ implies
\bea
\Lambda_{\un A \ov B} = \Lambda_{\ov A \un B} = 0 \, .
\eea

The generalized dilaton is a double Lorentz invariant, and its transformation under generalized diffeomorphisms is not covariant. However $e^{-2d}$ transforms as a generalized scalar density with $\omega=1$. The generalized frame transforms as a vector under double Lorentz transformations, and therefore,
\bea
\delta_{\Lambda} k_{\dot \mu \underline{\dot a}} = k_{\dot \mu \underline{\dot b}} \Lambda^{\underline{\dot b}}{}_{\underline{{\dot a}}}, \quad \delta_{\Lambda} \bar k_{\dot \mu \overline{\dot a}} = \bar k_{\dot \mu \overline{\dot b}} \Lambda^{\overline{\dot b}}{}_{\overline{\dot a}} \, , \\
\delta_{\Lambda} h^{\dot \mu}{}_{\underline{\dot a}} = h^{\dot \mu}{}_{\underline{\dot b}} \Lambda^{\underline{\dot b}}{}_{\underline{{\dot a}}}, \quad \delta_{\Lambda} \bar h^{\dot \mu}{}_{\overline{\dot a}} = \bar h^{\dot \mu}{}_{\overline{\dot b}} \Lambda^{\overline{\dot b}}{}_{\overline{\dot a}} \, .
\eea

The previous transformations will define a bigravity structure since we will use both $(k,\bar K)$ and $(h,\bar h)$ to create the same non-relativistic degrees of freedom. A similar scenario, but for a relativistic bigravity theory coming from string theory was studied in \cite{Bi}.

Acting on a generic vector the Lorentz derivative is defined as
\be
\nabla_{A}V_{B}= E_{A}V_{B} + \omega_{AB}{}^{C}V_{C}\, 
\ee
where $E_{A}\equiv \sqrt2{E}_{A}{}^{M}{}\partial_{M}$ and $\omega_{AB}{}^{C}$ is a spin connection which satisfies
\be
\omega_{ABC} = - \omega_{ACB}\, \qquad \mathrm{and} \qquad \omega_{A\overline B \underline C} = \omega_{A\underline B \overline C} = 0 \, .
\ee
Only the totally antisymmetric  and trace parts of $\omega_{ABC}$ can be determined in terms of the fundamental fields of the theory,  namely
 \bea
 \omega_{[{ABC}]} & = & -\frac13{ F}_{{ABC}}\, ,
\label{gralspinconnectionE}\\
\omega_{{B A}}{}^{ B} &=& -\sqrt{2} e^{2d} \partial_{ M}\left(E^{ M}{}_{ A} e^{-2d}\right)  \equiv -{ F}_{ A}\, ,
\label{gralspinconnectiontrace}
\eea
the latter arising from partial integration with the dilaton density 
\be
\int e^{-2d}V\nabla_{ A}V^{ A}=-\int  e^{-2d}V^{ A}\nabla_{A}V\, ,
\ee
for arbitrary $V$ and $V^{A}$. The relevant components of the generalized fluxes for computing the bosonic and fermionic heterotic DFT action are 
\bea
\Big\{ F_{\un A}, F_{\un A \un B \un C}, F_{\ov A \un B \un C}, F_{\un A \ov B \ov C} \Big\} \, .
\label{Relevant}
\eea 

The parametrization of the useful components for describing the bosonic sector of the theory is given by
 \begin{subequations}\label{fluxes}
\begin{align}
F_{\overline{\dot a}\underline{ {\dot b}{\dot c}}}&= -\frac12(\bar h^{\dot \nu}{}_{\ov {\dot a}} \partial_{\dot \nu} h^{\dot \mu}{}_{\un{\dot b}} k_{\dot \mu \un{\dot c}} + \bar h^{\dot \nu}{}_{\ov{\dot a}} \partial_{\dot \nu} k_{\dot \mu \un{\dot b}} h^{\dot \mu}{}_{\un {\dot c}})  + h^{\dot \nu}{}_{[\un{\dot b}} \partial_{\dot \nu} h^{\dot \mu}{}_{\un{\dot c}]} \bar k_{\dot \mu \ov{\dot a}} \nn \\ & \quad - h^{\dot \nu}{}_{[\un{\dot b}} \partial_{\dot \nu} k_{\dot \mu \un{\dot c}]} \bar h^{\dot \mu}{}_{\ov{\dot a}} - \frac12 \bar h^{\dot \mu}{}_{\ov{\dot a}} h^{\dot \nu}{}_{\un{\dot b}} h^{\dot \rho}{}_{\un{\dot c}} H_{\dot \mu \dot \nu \dot \rho}\, ,\\
F_{\underline{\dot a} \overline{{\dot b}{\dot c}}}& = \frac12( h^{\dot \nu}{}_{\un{\dot a}} \partial_{\dot \nu} \bar h^{\dot \mu}{}_{\ov{\dot b}} \bar k_{\dot \mu \ov{\dot c}} +  h^{\dot \nu}{}_{\un{\dot a}} \partial_{\dot \nu} \bar k_{\dot \mu \ov{\dot b}} \bar h^{\dot \mu}{}_{\ov{\dot c}})  - \bar h^{\dot \nu}{}_{[\ov{\dot b}} \partial_{\dot \nu} \bar h^{\dot \mu}{}_{\ov{\dot c}]} k_{\dot \mu \un{\dot a}} \nn \\ & \quad + \bar h^{\dot \nu}{}_{[\ov{\dot b}} \partial_{\dot \nu} \bar k_{\dot \mu \ov{\dot c}]}  h^{\dot \mu}{}_{\un{\dot a}} -\frac12 h^{\dot \mu}{}_{\un{\dot a}} \bar h^{\dot \nu}{}_{\ov{\dot b}} \bar h^{\dot \rho}{}_{\ov{\dot c}} H_{\dot \mu \dot \nu \dot \rho}\, ,\\
F_{\underline{{\dot a} {\dot b} {\dot c}}}& = -\frac32( h^{\dot \nu}{}_{[\un{\dot a}} \partial_{\dot \nu} h^{\dot \mu}{}_{\un{\dot b}} k_{\dot \mu \un{\dot c}]} + h^{\dot \nu}{}_{[\un{\dot a}} \partial_{\dot \nu} k_{\dot \mu \un{\dot b}} h^{\dot \mu}{}_{\un{\dot c}]})- \frac12 h^{\dot \mu}{}_{\un{\dot a}} h^{\dot \nu}{}_{\dot{\un b}} h^{\dot \rho}{}_{\un{\dot c}} H_{\dot \mu \dot \nu \dot \rho}\, ,\\
F_{\un{\dot a}} & = \partial_{\dot \mu}h^{\dot \mu}{}_{\un{\dot a}} - 2 \partial_{\un{\dot a}}(\phi - \frac12 ln(e))\, ,
\end{align}
 \end{subequations}
where the H-flux is given by $H_{\dot \mu \dot \nu \dot \rho}=3 \partial_{[\dot \mu}B_{\dot \nu \dot \rho]}$.

\item {\bf Action principle:}
\end{enumerate}
The DFT action is given by
\bea
\int d^{2D}X e^{-2d} {\cal R} \, ,
\eea
where ${\cal R}(E,d)$ is a two derivative scalar under generalized diffeomorphisms and it is invariant under Lorentz transformations. This object is known as the generalized Ricci scalar, and can be written in terms of the generalized fluxes \cite{FF},
\bea
{\cal R}& = & 2E_{\un{A}}F^{\un{A}} + F_{\un{A}}F^{\un{A}} - \frac16{F}_{\un{ABC}}F^{\un{ABC}} - \frac12{F}_{\ov{A}\un{BC}}F^{\ov{A}\un{BC}} \, ,
\label{GR}
\eea

The full parametrization of the DFT action in terms of $H^{\dot \mu \nu}$, $K_{\dot \mu \dot \nu}$, their kernels, $B_{\dot \mu \dot \nu}$ and $\phi$ has been detailed study in \cite{NRaction} for torsional Newton-Cartan theory, Carrollian theory and string Newton-Cartan theory. The analysis was performed considering the generalized metric formalism of DFT, which is equivalent to the flux formalism. In this work we will compute the extra bosonic terms that appear in the heterotic extension of this theory, as well as the fermionic contributions to leading order in fermions. 

In this work we are focused in obtaining the minimal ${\cal N}=1$ supersymmetric extension for a general non-Riemannian DFT which means that our goul is to obtain a minimal $D=10$ supergravity defined on a non-Riemannian background. However, in principle, the inclusion of a Yang-Mills field in the parametrization of the generalized frame it is possible, for example,
\be
E^{M}{}_{ A}  \ = \
\frac{1}{\sqrt{2}}\left(\begin{matrix}- k_{\dot \mu \un a}-C_{\dot \rho \dot \mu} {h}^{\dot \rho }{}_{\un{\dot a}} &  h^{\dot \mu }{}_{\un {\dot a}} & -A_{\dot \rho}{}^\alpha {h}^{\dot \rho }{}_{\un{\dot a}}\, , \\ 
\bar k_{\dot \mu \ov{\dot a}}-C_{\dot \rho \dot \mu}{} \bar h^{\dot \rho }{}_{\ov {\dot a}}& \bar h^{\dot \mu}{}_{\ov{\dot a}}&-A_{\dot \rho}{}^\alpha  \bar{h}^{\dot \rho}{}_{\ov{\dot a}} \\
\sqrt{2} A_{\dot \mu \beta}e^\beta{}_{\overline \alpha} &0&\sqrt{2} e^\alpha{}_{\overline \alpha} \end{matrix}\right)  \, ,
\ee
where $C_{\dot \mu \dot \nu}=B_{\dot \mu \dot \nu} + \frac12 A_{\dot \mu}{}^{\alpha} \kappa_{\alpha \beta} A_{\dot \nu}{}^{\beta}$ mimicking \cite{MS} and $\kappa^{\alpha \beta}$ is a Cartan-Killing (inverse) metric which satisfies 
\bea
e^{\alpha}{}_{\ov \alpha} \kappa^{\ov \alpha \ov \beta} e^{\beta}{}_{\ov \beta}=\kappa^{\alpha \beta} \, .
\label{framewithA}
\eea
As usual the indices $\ov A$ have to decomposed as $=(\ov a, \ov \alpha)$ and we have to use $e^{\alpha}{}_{\ov \alpha}$ as a constant object to relate the Cartan-Killing metrics. Using this enlargement of the global symmetry from $O(D,D)$ to $O(D,D+n)$ with $n=496$ and considering the previous generalize frame it is now possible to construct the generalized metric for Heterotic DFT in non-Riemannian backgrounds as
\bea
{\cal H}^{\dot \mu \dot \nu} & = & H^{\dot \mu \dot \nu} \nn \\
{\cal H}^{\dot \mu}{}_{\dot \nu} & = & - C_{\dot \rho \dot \nu} H^{\dot \rho \dot \mu} - x_{\dot \nu} y^{\dot \mu} + \bar x_{\dot \nu} \bar y^{\dot \mu}  \nn \\
{\cal H}^{\dot \mu}{}_{\alpha} & = & - H^{\dot \mu \dot \rho} A_{\dot \rho \alpha} \nn \\
{\cal H}_{\dot \mu \dot \nu} & = & K_{\dot \mu \dot \nu} - C_{\dot \mu \dot \sigma} H^{\dot \sigma \dot \rho} C_{\dot \rho \dot \nu} - 2 x_{(\dot \mu} C_{\dot \nu) \dot \rho} y^{\dot \rho} + 2 \bar x_{(\dot \mu} C_{\dot \nu) \dot \rho} \bar y^{\dot \rho} - A_{\dot \mu \alpha} A_{\dot \nu}{}^{\alpha} \nn \\
{\cal H}_{\dot \mu \alpha} & = & A_{\dot \rho \alpha} (C_{\dot \sigma \dot \mu} H^{\dot \sigma \dot \rho} +\delta_{\dot \mu}^{\dot \rho} + x_{\dot \mu} y^{\dot \rho} - \bar x_{\dot \mu} \bar y^{\dot \rho}) \nn \\
{\cal H}_{\alpha \beta} & = & \kappa_{\alpha \beta} + A_{\dot \rho \alpha} H^{\dot \rho \dot \sigma} A_{\dot \sigma \beta} \, .
\eea
This metric can be used \footnote{See also \cite{DYM} for formulations with explicit double Yang Mills fields in the double geometry.} to include Yang-Mills contributions to the bosonic sector of the theory, while the generalized frame given in (\ref{framewithA}) is the one to construct the full heterotic version of this model. In this work we will keep $A_{\mu}=0$ for simplicity.

\section{${\cal N}=1$ supersymmetric extension for a general non-Riemannian Double Field Theory}

\subsection{General construction}

The ${\cal N}=1$ supersymmetric extension of DFT is achieved by fixing the dimension of the target space ($D=10$) and adding generalized spinor fields that act as supersymmetric partners of the bosonic fields: the generalized gravitino $\Psi_{\ov{A}}$ and the generalized dilatino $\rho$. 

The covariant derivative of spinor fields acquires an additional term in order to derive the spinor indices. For instance, 
\bea
\nabla_{A}\Psi_{\ov{B}} & = & E_{A}\Psi_{\ov{B}} + \omega_{A\ov{B}}{}^{\ov{C}}\Psi_{\ov{C}} - \frac{1}{4}\omega_{A\un{BC}}\gamma^{\un{BC}}\Psi_{\ov{B}}\, , \nn\\
\nabla_{A}\rho & = & E_{A}\rho - \frac{1}{4} \omega_{A\un{BC}}\gamma^{\un{BC}}\rho\,  . 
\eea
The gamma matrices satisfy a Clifford algebra,
\be
\left\{\gamma^{\un{A}},\gamma^{\un{B}}\right\} = - 2P^{\un{AB}}\ , \label{Cliff}
\ee
and we use the standard convention for antisymmetrization of $\gamma$-matrices $\gamma^{{\underline A \dots \underline B}}=\gamma^{[\underline{A}} \dots \gamma^{\underline{B}]}$.  

The generalized supersymmetry transformations of the fundamental fields are parameterized by an infinitesimal Majorana fermion $\epsilon$, that is a spinor of $O(1,9)_L$,   
\bea
\label{0transfA}
\delta E^{M}{}_{\un{A}}{} & = &  - \frac12 \ov{\epsilon}\gamma_{\un{A}} \Psi^{\ov{B}}E^{M}{}_{\ov{B}}\, , \\   
\delta E^{M}{}_{\ov{A}}{} & = & \frac12 \ov{\epsilon}\gamma^{\un{B}}\Psi_{\ov{A}}E^{M}{}_{\un{B}}{}\, ,\\   
\delta d & = &  - \frac{1}{4}\ov{\epsilon}\rho \, , \label{0transfB}   
\eea
from where we read
\bea
\delta h^{\dot \mu \underline{\dot a}} & = & - \frac12 \bar \epsilon \gamma^{\underline{\dot a}} \psi^{\overline{\dot b}} \bar h^{\dot \mu}{}_{\overline{\dot b}}
 \, ,\quad
\delta \bar h^{\dot \mu \overline{\dot a}} =  \frac12 \bar \epsilon \gamma^{\underline{\dot b}} \psi^{\overline{\dot a}} h^{\dot \mu}{}_{\underline{\dot{b}}} \\
\delta k_{\dot \mu \underline{\dot a}} 
& = & - \frac12 \bar \epsilon \gamma^{\underline{\dot c}} k_{\dot \mu \underline{\dot c}} \psi^{\overline{\dot b}} e_{\underline{\dot a} \overline{\dot b}} \quad \delta \bar k_{\dot \mu \underline {\dot a}} =  \frac12 \bar \epsilon \gamma^{\underline{\dot a}} e_{\overline{\dot a} \underline{\dot a}} \psi^{\overline{\dot b}} \bar k_{\dot \mu \overline{\dot b}} \, ,
\eea
with $e_{\underline{\dot a} \overline{\dot b}}= - e_{\overline{\dot b} \underline{\dot a}} = h^{\dot \mu}{}_{\underline{\dot a}} \bar k_{\dot \mu \overline{\dot b}}$. Interestingly enough, $k_{\dot \mu \dot a}$ and $\bar k_{\dot \mu \dot a}$ cannot be identified using a gauge fixing procedure unless $x_{\dot \mu}=\bar x_{\dot \mu}=y^{\dot \mu}=\bar y^{\dot \mu}=0$, and the same happens with $h^{\dot \mu}{}_{\underline{\dot a}}$ and $\bar h^{\dot \mu}{}_{\overline{\dot a}}$ and the double Lorentz parameters. This is a very interesting point of the construction since, typically, the ${\cal N}=1$ supersymmetric DFT requires that one of the double parameters is related to the other up to fermionic contributions (gauge fixing condition). In these kind of (non-Riemannian) geometries, some of the supersymmetric transformations depend on the products $k_{\dot \mu \un{\dot a}} \ov h^{\dot \mu}{}_{\ov{\dot b}}$ or $\ov k_{\dot \mu \ov{\dot a}} \un h^{\dot \mu}{}_{\un{\dot b}}$ which are typically Minkowski metrics in Riemannian geometries due to the identifications of $h\leftrightarrow \bar h$ and $k \leftrightarrow \bar k$ in those cases. Although this is not a problem for the minimal supergravity model, the inclusion of the Yang-Mills sector coming from a Heterotic DFT could be not straightforward. We will return to this point in the final section.    

On the other hand, the supersymmetric transformation for the B-field is given by
\bea
\delta B_{\dot \mu \dot \nu} = \bar \epsilon \gamma^{\underline{\dot a}} k_{[\dot \mu \underline{\dot a}} \psi^{\overline{\dot b}} \bar k_{\dot \nu] \overline{\dot b}} \, ,
\eea
while the transformations for the fermionic degrees of freedom take the usual form,
\bea
\delta\Psi_{\ov{\dot a}} & = &  \nabla_{\ov{\dot a}}\epsilon \, ,\quad
\delta\rho =  - \gamma^{\un{\dot a}}\nabla_{\un{\dot a}}\epsilon \, .
\eea

The symmetry transformation rules leave the following action invariant,
\bea
\label{DFTsusyAction}
S_{\mathcal{N}=1 \ \textrm{DFT}} & = & \int d^{20}X e^{-2d} \Big( {\cal R} + \ov{\Psi}^{\ov{A}}\gamma^{\un{B}}\nabla_{\un{B}}\Psi_{\ov{A}} - \ov{\rho} \gamma^{\un{A}}\nabla_{\un{A}}\rho + 2\ov{\Psi}^{\ov{A}}\nabla_{\ov{A}}\rho \Big) \, ,
\eea
where ${\cal R}={\cal R}(H,K,B,e,x,y)$ and $e^{-2d}=e$. The fermionic Lagrangian of the previous expression can be written in terms of the different components of (\ref{Relevant}) as
\bea
L_{\psi \psi} & = & \bar \Psi^{\ov{\dot a}} \gamma^{\un{\dot b}} E_{\un {\dot b}} \Psi_{\ov{\dot a}} - \bar \Psi^{\ov{\dot b}} \gamma^{\un{\dot a}} (\frac12( h^{\dot \nu}{}_{\un{\dot a}} \partial_{\dot \nu} \bar h^{\dot \mu}{}_{\ov{\dot b}} \bar k_{\dot \mu \ov{\dot c}} +  h^{\dot \nu}{}_{\un{\dot a}} \partial_{\dot \nu} \bar k_{\dot \mu \ov{\dot b}} \bar h^{\dot \mu}{}_{\ov{\dot c}})  - \bar h^{\dot \nu}{}_{\ov{\dot b}} \partial_{\dot \nu} \bar h^{\dot \mu}{}_{\ov{\dot c}} k_{\dot \mu \un{\dot a}} \nn \\ && + \bar h^{\dot \nu}{}_{\ov{\dot b}} \partial_{\dot \nu} \bar k_{\dot \mu \ov{\dot c}}  h^{\dot \mu}{}_{\un{\dot a}} -\frac12 h^{\dot \mu}{}_{\un{\dot a}} \bar h^{\dot \nu}{}_{\ov{\dot b}} \bar h^{\dot \rho}{}_{\ov{\dot c}} H_{\dot \mu \dot \nu \dot \rho}) \Psi^{\ov{\dot c}} \nn \\ && - \bar \Psi^{\ov{\dot d}} \gamma^{\un{\dot a} \un{\dot b} \un{\dot c}} (\frac18( h^{\dot \nu}{}_{\un{\dot a}} \partial_{\dot \nu} h^{\dot \mu}{}_{\un{\dot b}} k_{\dot \mu \un{\dot c}} + h^{\dot \nu}{}_{\un{\dot a}} \partial_{\dot \nu} k_{\dot \mu \un{\dot b}} h^{\dot \mu}{}_{\un{\dot c}}) + \frac{1}{24} h^{\dot \mu}{}_{\un{\dot a}} h^{\dot \nu}{}_{\dot{\un b}} h^{\dot \rho}{}_{\un{\dot c}} H_{\dot \mu \dot \nu \dot \rho}) \Psi_{\ov{\dot d}} \nn \\ && + \bar \Psi^{\ov{\dot a}} \gamma^{\un{\dot c}} (\frac12 \partial_{\dot \mu}h^{\dot \mu}{}_{\un{\dot c}} - \partial_{\un{\dot c}}(\phi - \frac12 ln(e))) \Psi_{\ov{\dot a}}  \, ,
\eea
\bea
L_{\psi \rho} & = &  2 \bar \Psi^{\ov{\dot a}} E_{\ov{\dot a}} \rho + \frac12 \bar \Psi^{\ov{\dot a}} \Big[ -\frac12(\bar h^{\dot \nu}{}_{\ov {\dot a}} \partial_{\dot \nu} h^{\dot \mu}{}_{\un{\dot b}} k_{\dot \mu \un{\dot c}}  + \bar h^{\dot \nu}{}_{\ov{\dot a}} \partial_{\dot \nu} k_{\dot \mu \un{\dot b}} h^{\dot \mu}{}_{\un {\dot c}}) \nn \\ && + h^{\dot \nu}{}_{\un{\dot b}} \partial_{\dot \nu} h^{\dot \mu}{}_{\un{\dot c}} \bar k_{\dot \mu \ov{\dot a}}  - h^{\dot \nu}{}_{\un{\dot b}} \partial_{\dot \nu} k_{\dot \mu \un{\dot c}} \bar h^{\dot \mu}{}_{\ov{\dot a}} - \frac12 \bar h^{\dot \mu}{}_{\ov{\dot a}} h^{\dot \nu}{}_{\un{\dot b}} h^{\dot \rho}{}_{\un{\dot c}} H_{\dot \mu \dot \nu \dot \rho} \Big] \gamma^{\un{\dot b} \un{\dot c}} \rho \, ,
\eea
\bea
L_{\rho \rho} & = &  - \bar \rho \gamma^{\un{\dot a}} E_{\un{\dot a}} \rho + \bar \rho \gamma^{\un{\dot a} \un{\dot b} \un{\dot c}} (\frac{1}{8}( h^{\dot \nu}{}_{\un{\dot a}} \partial_{\dot \nu} h^{\dot \mu}{}_{\un{\dot b}} k_{\dot \mu \un{\dot c}} + h^{\dot \nu}{}_{\un{\dot a}} \partial_{\dot \nu} k_{\dot \mu \un{\dot b}} h^{\dot \mu}{}_{\un{\dot c}}) + \frac{1}{24} h^{\dot \mu}{}_{\un{\dot a}} h^{\dot \nu}{}_{\dot{\un b}} h^{\dot \rho}{}_{\un{\dot c}} H_{\dot \mu \dot \nu \dot \rho}) \rho \nn \\ && - \bar \rho \gamma^{\un{\dot b}} (\frac12 \partial_{\dot \mu}h^{\dot \mu}{}_{\un{\dot b}} - \partial_{\un{\dot b}}(\phi - \frac12 ln(e))) \rho  \, .
\eea

So far this is the most general form of the leading order in fermions Lagrangian coming from the non-relativistic parametrization of ${\cal N}=1$ DFT. In here we are focusing in the minimal D=10 supergravity model which means that we are not taking into account the contributions coming from the gauge field $A$ and its supersymmetric partner the gaugino. While the former should be encoded through the generalized frame the latter should be part of the generalized gravitino as happens in the Riemannian cases. In the next subsection we will show the ${\cal N}=1$ supersymmetric extension of type I TNC geometry, which corresponds to a $(n,\bar n)=(1,1)$ non-riemannian DFT using the classification given in \cite{PM}.

\subsection{${\cal N}=1$ supersymmetric type I TNC geometry}

TNC geometries are characterized by local Galilean symmetry. When coming from the flux formulation of DFT we need to think about this theory as a bymetric gravity with a double tangent  space structure. This is not the case in \cite{NRaction} since the generalized metric formulation is enough to describe the bosonic sector of the theory. 

The non-relativistic spacetime is embedded using a pair of spatial frames  $k^{\un a}_\mu$ and $k^{\ov a}_\mu$ used to define a spatial (or transverse) metric
\begin{align}
	h_{\mu \nu} &= k_\mu{}_{\un a} k_\nu{}^{\un a} = \ov k_\mu{}_{\ov a} \ov k_\nu{}^{\ov a} \, ,
\end{align}
 and a temporal frame $\tau_\mu$ used as a universal one-form clock. We also introduce a one-form masses $m_\mu$ in order to have the following bosonic symmetry transformations, 
\begin{align}
 	\begin{split}\label{transfFrames}
   \delta \ov k_\mu{}^{\ov a} &= \mathcal{L}_\xi \ov k_\mu{}^{\ov a} + \Lambda^{\ov a} \tau_\mu + \Lambda^{\ov a}{}_{\ov b} \ov k_\mu{}^{\ov b} \,, \\
 		\delta \tau_\mu &= \mathcal{L}_\xi \tau_\mu \, , \\
 		\delta m_\mu &= \mathcal{L}_\xi m_\mu + \Lambda_{\un a} k_\mu{}^{\un a} + \partial_\mu \sigma \, ,
 	\end{split}
 \end{align} 
with $\Lambda^{\un a}$ and $\Lambda^{\ov a}$ local Galilean boost parameters which satisfy $\Lambda^{\un a} k_{\mu \un a}=\Lambda^{\ov a} \ov k_{\mu \ov a}$ and $\sigma$ is a $U(1)$ gauge transformation parameter. The square matrices $(\tau_\mu,k_\mu{}^{\un a})$ and $(\tau_\mu,\ov k_\mu{}^{\ov a})$ have inverses $(-\upsilon^\mu, h^\mu{}_{\un a})$ and $(-\upsilon^\mu, \ov h^\mu{}_{\ov a})$ where these inverse frames are orthogonal
\be
		\upsilon^\mu \tau_\mu = -1 \, , \qquad
		k^\mu{}_{\un a} \tau_\mu = \ov k^\mu{}_{\ov a} \tau_\mu = 0 \, , \qquad
		h_\mu{}^{\un a} \upsilon^\mu = \ov h_\mu{}^{\ov a} \upsilon^\mu = 0 \, ,
\ee
and complete
 \be\label{completFrames}
  k_\nu{}^{\un a} k^\mu{}_{\un a} - \upsilon^\mu \tau_\nu =\delta ^\mu_\nu	\, ,  \quad \ov k_\nu{}^{\ov a} \ov k^\mu{}_{\ov a} - \upsilon^\mu \tau_\nu =\delta ^\mu_\nu
 \ee
Making use of inverse spatial frames $h^\mu{}_{\un a}$ and $\ov h^\mu{}_{\ov a}$ one can define an inverse spatial metric as 
\begin{align}
	h^{\mu \nu} &= h^\mu{}_{\un a} h^\nu{}_{\un a} = \ov h^\mu{}_{\ov a} \ov h^\nu{}_{\ov a} \, .
\end{align}

From the DFT generalized frame we obtain the following relations,
\bea
- k_{\mu}{}^{\un a} k_{\nu \un a} & = & \ov k_{\mu}{}^{\ov a} \ov k_{\nu \ov a} = h_{\mu \nu} \\
- h^{\mu}{}_{\un a} h^{\nu \un a} & = & \ov h^{\mu}{}_{\ov a} \ov h^{\nu \ov a} = h^{\mu \nu} \\
- h^{\mu}{}_{\un a} h^{\un a} & = & \ov h^{\mu}{}_{\ov a} \ov h^{\ov a} = h^{\mu \rho} \aleph_{\rho} \, ,
\eea
and we set $e^{-2d}=e$, while the B-field and the kernels are given by
\begin{align}\label{TNCDFT1}
{B}_{\dot \mu \dot \nu} &=  \begin{pmatrix}
			\bar B_{\mu\nu} && -m_\mu \\ 
			m_\nu && 0
		\end{pmatrix}
\end{align}
and
\begin{align} 
x_{\dot \mu}&=\frac{1}{\sqrt{2}}  \begin{pmatrix}
			\tau_\mu - \aleph_\mu  \\ 
			 1 
		\end{pmatrix},&
\bar x_{\dot \mu}& =\frac{1}{\sqrt{2}}  \begin{pmatrix}
			\tau_\mu+ \aleph_\mu  \\ 
			 -1 
		\end{pmatrix},\nonumber\\
y^{\dot \mu}& =\frac{1}{\sqrt{2}}  \begin{pmatrix}
			-\upsilon^\mu  \\ 
			 1 -\upsilon^\mu \aleph_\mu
		\end{pmatrix},&
\bar y^{\dot \mu} &=\frac{1}{\sqrt{2}}  \begin{pmatrix}
			-\upsilon^\mu  \\ 
			 -1-\upsilon^\mu \aleph_\mu
		\end{pmatrix}. \label{TNCDFT2}
\end{align}
We also define
\be
b_{\mu\nu} \equiv \partial_{[\mu}\aleph_{\nu]}, \qquad \qquad \mathfrak{e}_\mu \equiv \hat 
\upsilon^\rho b_{\rho\mu} \, .
\ee

In order to read the sypersymmetric extension to the transformations rules we decompose the generalized gravitino in the following way
\bea
\Psi_{\ov{A}} = \Psi_{\ov{\dot a}} = (\Psi_{\ov{a}}, \psi , \tilde \psi) \, ,
\eea
while the DFT gamma matrices as $\gamma_{\underline{\dot a}}= (\gamma_{\underline a},1,i)$. Furthermore, we keep $\rho$ as the $10$-dimensional dilatino field. The supersymmetric extension for the transformations is therefore given by
\bea
\delta \ov h^{\mu}{}_{\ov a} & = & \frac12 \bar \epsilon \gamma^{\nu} \psi_{\ov a} (\delta^{\mu}_{\nu} - \aleph_{\nu} v^{\mu}) \\
\delta v^{\mu} & = & \frac{1}{\sqrt 2} \bar \epsilon \psi^{\nu} (\delta^{\mu}_{\nu} + \aleph_{\nu} v^{\mu}) \, 
\eea
while the universal clock transforms as
\bea
\delta \tau_{\mu} = \frac{1}{\sqrt2} \tau_{\mu} \bar \epsilon \psi^{\nu} (\tau_{\nu} - \aleph_{\nu}) \, .
\eea
Finally the transformation of the B-field is given by
\bea
\delta B_{\mu \nu} = \bar \epsilon \gamma^{\underline{\dot a}} k_{[\mu \underline{\dot a}} \psi^{\overline{\dot b}} \bar k_{\nu] \overline{\dot b}} \, .
\eea
The present model constitutes a convenient example to show how the ${\cal N}=1$ supersymmetric extension of the non-Riemannian DFT works. Since this geometry is related to Carrollian geometry and string Newton Cartan throught particular T-duality rotations or a null reduction/uplift, the formalism here introduced applies also in that scenarios. In the next part of the work we discuss about potential uses of the present model. 

\section{Future directions}
In this work we have explored non-Riemannian geometries using the flux formalism of DFT and we include fermionic degrees of freedom in order to construct a ${\cal N}=1$ supersymmetric invariant formulation. From the supersymmetric transformations rules we have extract the transformations for the non-Riemannian parametrization of the generalized frame. We have constructed the generalized fluxes required for the bosonic and fermionic action and we gave the most general form of the action in order to capture the supersymmetric extension of the non-Riemannian geometries with $n=\bar n$. As an example we show the supersymmetric extension for type I TNC, which corresponds to a particular case where $n=\bar n=1$. We finish this work with some possible continuations for this project  

\begin{itemize}
\item Here we have explored the exact form of the ${\cal N}=1$ supersymmetric DFT compatible with non-Riemannian exact backgrounds. It would be interesting to consider perturbations around these generic backgrounds starting from the DFT framework, i.e., to perturbe the generalized frame, dilaton, gravitino and dilatino (the last two with possible applications to fermionic condensates). As a simple model it would be nice to perturbe the aforementioned DFT background fields using the generalized Kerr-Schild ansatz which is a linear perturbation in the generalized frame using a pair of generalized null vectors \cite{hetKS}. Using this model one would be able to extract the perturbative form of the degrees of freedom for the non-Riemannian supergravity and explore notions of single/zeroth copy for these geometries.   

\item We restrict our supersymmetric extension to minimal non-Riemannian supergravity models which can be obtained from DFT. The inclusion of the Yang-Mills sector is another possible continuation, which is straightforward for type I TNC since in this work we gave the prescription for the generalized frame. However, in more general scenarios, like type II TNC, it would be necessary to generalized the method given in \cite{type2}, take the relativistic limit in terms DFT fields, and then to take advantage of the dependence of the generalized frame on the gauge field (and the generalized gravitino on the gaugino) in order to read the vectorial part of the model. It would be very interesting to recover this gauge sector of heterotic supergravity defined on a non-Riemannian geometry from a DFT model. 

\item The inclusion of fermionic degrees of freedom for non-Riemannian geometries coming from a DFT formulation opens the possibility of non-Riemannian extensions for the superspace formulation of DFT \cite{superDFT}, where a suitable generalization for the generalized super frame could be the way of regroup all the degrees of freedom of the theory making use of the orthosymplectic extension of the $O(D,D)$ group. This procedure could give some insights about how to obtain the supersymmetric sector of Heterotic supergravity coming from a non-Riemannian formulation \cite{KK}.

\end{itemize}

\subsection*{Acknowledgements}
We thank Sourav Roychowdhury for interesting discussions and Jan Rosseel for enlightening comments and discussions. We thank IPhT-CEA Saclay for hospitality during the last stage of this project. We also thank to the French-Croatian bilateral project “Physics of non-geometric fluxes” of the Cogito programme 2021-2022 and,  particularly, to Mariana Grana and Athanasios Chatzistavrakidis. E.L is supported by the Croatian Science Foundation project IP-2019-04-4168.

\end{document}